# pDSurfTomo: A High-Performance Parallel Computing Package for Direct Surface Wave Tomography

Shaohang Zhu[1,2], Junlun Li[1,2,3*], Guoyi Chen[1,2], Hongjian Fang[4], Huajian Yao[1,2,3]

[1]State Key Laboratory of Precision Geodesy, School of Earth and Space Sciences, University of Science and Technology of China, Hefei 230026, China

[2]Mengcheng National Geophysical Observatory, University of Science and Technology of China, Mengcheng 233500, China

[3]Anhui Provincial Key Laboratory of Subsurface Exploration and Earthquake Hazard Risk Prevention, Hefei 230031, China

[4]School of Earth Sciences and Engineering, Sun Yat-sen University, Zhuhai, China

*Corresponding author: lijunlun@ustc.edu.cn

# Abstract

Surface wave tomography is essential for investigating the shear-wave velocity structure of the crust and upper mantle. The direct surface wave tomography method, DSurfTomo, has become one of the most widely adopted surface wave tomography packages due to its ability to account for ray path bending in complex media to increase subsurface characterization accuracy. However, the inherent serial architecture of DSurfTomo lacks effective support for multicore CPU and GPU parallel computing. Furthermore, its built-in solver is computationally expensive when solving large-scale linear systems. Consequently, the software struggles to meet current demands for large-scale, high-resolution surface wave tomography. To address these limitations, we propose pDSurfTomo, a highly optimized, next-generation software package utilizing hybrid CPU-GPU acceleration. First, it overcomes the scalability bottleneck associated with the sensitivity kernel computation through a refined parallel design; also, it uses vectorization techniques to accelerate the forward modeling of surface wave dispersion, achieving efficient computation of the sensitivity kernel. Second, it implements parallelization of the serial fast marching method using OpenMP, significantly reducing the computation time for theoretical surface wave traveltimes. Finally, it incorporates GPU acceleration to efficiently solve large-scale sparse linear least-squares problems. To further streamline the workflow, we provide a cross-platform graphical user interface with remote server connectivity, which allows users to execute and visualize inversion tasks locally while seamlessly utilizing remote computing clusters. Application to an observed dispersion dataset collected from 229 stations in North China demonstrates that pDSurfTomo reduces the computation time by more than an order of magnitude while maintaining a negligible discrepancy compared to the original DSurfTomo. It is expected that pDSurfTomo will provide a highly efficient and accessible solution for large-scale, high-resolution surface wave tomography.

# Introduction

Surface wave tomography plays a crucial role in resolving seismic velocity and anisotropy in the crust and upper mantle (e.g., Woodhouse and Dziewonski, 1984; Yao et al., 2010; Fang et al., 2015; Liu et al., 2019; Hu et al., 2020; Deng et al., 2022). As surface waves propagate, depth-dependent variations in the Earth's elastic properties cause waves of different frequencies to travel at distinct speeds, a phenomenon known as dispersion. This dispersive characteristic allows the phase or group velocities of surface waves at different frequencies to effectively constrain the velocity structures at varying depths within the Earth.

Conventional surface wave tomography based on dispersion data usually has two steps. First, it inverts for 2D phase/group velocity maps (e.g., Trampert and Woodhouse, 1995; Yao et al., 2006; Chen et al., 2026); then, it conducts a point-wise inversion to obtain a 1D shear-wave velocity model at each grid point, which are subsequently combined to construct a 3D velocity model (Shapiro and Ritzwoller, 2002; Yao et al., 2008). This method usually assumes that surface waves propagate along great circle paths, thereby neglecting bending of ray trajectories due to structural anomalies. Previous studies have demonstrated that when the medium exhibits strong lateral heterogeneity, traveltime anomalies induced by ray path bending cannot be ignored and must be thoroughly accounted for during the inversion process (Yang and Hung, 2005; Li et al., 2015). Furthermore, due to the sparse spatial distribution of surface wave paths at certain periods, it is difficult to construct reliable phase/group velocity maps for these periods. Consequently, this portion of the data cannot be properly leveraged in subsequent inversion for the 3-D shear-wave velocity model, thereby reducing the overall utilization of the dispersion data (Yao et al., 2022).

To circumvent the limitations of the two-step surface wave tomography, an alternative one-step approach has been proposed, which directly inverts the 3-D shear-wave velocity models using all observed dispersion traveltime data across various paths (Boschi and Ekström, 2002; Feng and An, 2010; Fang et al., 2015). A representative example is the direct surface wave tomography method (DSurfTomo) and its associated software package by Fang et al. (2015). This method accounts for the ray path bending effects at different periods by using the fast marching method (FMM) to conduct ray tracing at each period (Rawlinson and Sambridge, 2003), and obtains more accurate surface wave traveltimes. Furthermore, this approach incorporates dispersion traveltime data from all available periods directly into the inversion process, which significantly increases data utilization and ray path density for more reliable inversion results (Li et al., 2022; Li et al., 2023). Currently, DSurfTomo has been extensively applied in investigating shear-wave velocity structures across various scales, establishing itself as one of the most widely used software packages in surface wave tomography (e.g., Chen et al., 2016; Zhou et al., 2019; Nimiya et al., 2020; Nthaba et al., 2022; Zhao et al., 2023; Li et al., 2024; Zhai et al., 2025).

However, the DSurfTomo package possesses inherent architectural limitations which restrict its applicability to large-scale, high-resolution surface wave tomography. On the one hand, the package predominantly relies on a serial code architecture. Although parallelization is implemented for certain computational tasks during the inversion, the degree of parallelism is low and accounts for only a minor fraction of the total inversion time, yielding limited improvement to the overall inversion efficiency. Constrained by this serial architecture, the computation time required for inversion increases sharply as the grid resolution and the volume of surface wave dispersion data scale up. On the other hand, the package lacks support for GPU-based heterogeneous computing architectures, preventing it from leveraging the massively parallel processing capabilities to improve computational efficiency.

To date, no systematic efforts have addressed the parallel optimization of the inherently serial architecture in DSurfTomo. Consequently, this widely adopted algorithm remains unable to fully leverage the computational capabilities of modern heterogeneous hardware architectures, severely restricting its computational efficiency. To enhance the inversion efficiency of DSurfTomo for large-scale surface wave tomography, we propose pDSurfTomo in this study. First, we parallelize the original serial computational tasks using OpenMP. By rationally partitioning computational tasks and implementing efficient thread synchronization mechanisms, we enable the parallel execution of previously serial computational tasks on multicore CPUs. Second, to address the inadequate degree of parallelization in existing parallel tasks, we introduce a finer-grained parallel design across three dimensions: decoupling computational models, decomposing computational tasks, and tuning memory access patterns. Finally, to efficiently solve large-scale sparse linear least-squares problems, we implement a hybrid CPU-GPU acceleration strategy, offloading computationally demanding operations such as sparse matrix-vector multiplication to GPU to leverage the massively parallel processing capabilities of streaming multiprocessors. Furthermore, to streamline the workflow of the inversion, we develop a cross-platform graphical user interface (GUI) based on Python and PyQt5, which has robust compatibility across Windows and Linux environments. A defining feature of this application is its seamless remote server connectivity: Users can configure inversion parameters, dispatch computing tasks, and visualize output models on a remote cluster entirely through the local client, effectively circumventing the need for direct command-line interaction. This frontend-backend architecture significantly lowers the technical barrier for large-scale inversions and enhances overall research efficiency.

## Theoretical Framework of DSurfTomo

To provide necessary context for the computational optimizations and architectural enhancements introduced in this study, we first briefly review the core theoretical framework underlying the original DSurfTomo algorithm. Based on the initial 3-D P-wave velocity, S-wave velocity, and density model, the theoretical surface wave traveltimes between stations at different frequencies can be obtained through forward modeling. The traveltime residual between the observed and theoretical values for path $i$ at frequency $\omega$ can be expressed as:

$$\delta t_i(\omega) = t_i^{\text{obs}}(\omega) - t_i(\omega) = \sum_{k=1}^{K} v_{ik} \delta \hat{S}_k(\omega) \approx -\sum_{k=1}^{K} v_{ik} \frac{\delta C_k(\omega)}{C_k^2(\omega)}, \tag{1}$$

where $t_i^{\text{obs}}(\omega)$ and $t_i(\omega)$ represent the observed and theoretical surface wave traveltimes on path $i$,

respectively; $\delta \hat{S}_k(\omega)$ is the slowness perturbation of the surface wave propagation at horizontal grid point $k$; $C_k(\omega)$ and $\delta C_k(\omega)$ denote the phase (or group) velocity and the corresponding velocity perturbation at horizontal grid point $k$, respectively; and $v_{ik}$ is the bilinear interpolation coefficient on the $i$-th path.

The phase (or group) velocity perturbation $\delta C_k(\omega)$ can be expressed as:

$$\delta C_k(\omega) = \int \left[ \left. \frac{\partial C_k(\omega)}{\partial \alpha_k(z)} \right|_{\Theta_k} \delta \alpha_k(z) + \left. \frac{\partial C_k(\omega)}{\partial \beta_k(z)} \right|_{\Theta_k} \delta \beta_k(z) + \left. \frac{\partial C_k(\omega)}{\partial \rho_k(z)} \right|_{\Theta_k} \delta \rho_k(z) \right] dz, \quad (2)$$

where $\Theta_k$ denotes the 1-D layered velocity model in the depth direction at horizontal grid point $k$; $\alpha_k(z)$, $\beta_k(z)$, and $\rho_k(z)$ represent the P-wave velocity, S-wave velocity, and density at depth $z$ beneath horizontal grid point $k$, respectively. The sensitivity of the phase (or group) velocity to the model parameters ($\alpha$, $\beta$, $\rho$) at a given depth can be calculated via a simple finite-difference method. Following the empirical formulas proposed by Brocher (2005), the S-wave velocity can be converted into P-wave velocity and density:

$$\begin{aligned} \alpha(z) &= \sum_n \chi_n^{[\alpha]} \beta^n(z) \\ \rho(z) &= \sum_n \chi_n^{[\rho]} \beta^n(z) \end{aligned}. \quad (3)$$

Differentiating equation (3), we get:

$$\begin{aligned} \delta\alpha(z) &= \sum_n n\chi_n^{[\alpha]} \beta^{n-1}(z)\delta\beta(z) = R_\alpha(z)\delta\beta(z) \\ \delta\rho(z) &= \sum_n n\chi_n^{[\rho]} \beta^{n-1}(z)\delta\beta(z) = R_\rho(z)\delta\beta(z) \end{aligned}, \quad (4)$$

where $R_\alpha(z) = \sum_n n\chi_n^{[\alpha]} \beta^{n-1}(z)\delta\beta(z)$, $R_\rho(z) = \sum_n n\chi_n^{[\rho]} \beta^{n-1}(z)\delta\beta(z)$, $\chi_n^{[\alpha]}$ and $\chi_n^{[\rho]}$ are the coefficients of the fitting polynomials.

Combining equations (1), (2), and (4), we get:

$$\delta t_i(w) = \sum_{k=1}^{K}\left(-\frac{v_{ik}}{C_k^2(\omega)}\right)\sum_{j=1}^{J}\left[R_\alpha(z_j)\frac{\partial C_k(\omega)}{\partial \alpha_k(z_j)} + R_\rho(z_j)\frac{\partial C_k(\omega)}{\partial \rho_k(z_j)} + \frac{\partial C_k(\omega)}{\partial \beta_k(z_j)}\right]\bigg|_{\Theta_k} \delta\beta_k(z_j)$$
$$= \sum_{l=1}^{M} G_{il} m_l,\quad(5)$$

where $K$ and $J$ denote the number of grid points in the horizontal and vertical directions of the 3-D velocity model, respectively, with $M = KJ$ representing the total number of grid points in the 3-D velocity model. Consequently, solving for the model parameter $m_l$ can ultimately be transformed into solving the following system of linear equations:

$$\Delta d = G\Delta m, \quad (6)$$

where $\Delta d$ denotes the residual vector of the surface wave traveltime data across all paths and frequencies; $G$ represents the corresponding sensitivity kernel matrix of the surface wave traveltimes with respect to the 3-D S-wave velocity model, derived from the initial velocity model; and $\Delta m$ is the model perturbation vector of the 3-D S-wave velocity to be inverted:

$$\Delta m = [\delta\beta_1(z_1), \ldots, \delta\beta_1(z_J), \delta\beta_2(z_1), \ldots, \delta\beta_2(z_J), \delta\beta_K(z_1), \ldots, \delta\beta_K(z_J)]^T. \quad (7)$$

By solving this linear system via the least-squares method, we can obtain the 3-D S-wave velocity perturbation to update the S-wave velocity model. The corresponding P-wave velocity and density can then be calculated using the aforementioned empirical formulas (Brocher, 2005). Subsequently, the sensitivity kernel matrix $G$ is reconstructed using the updated 3-D velocity model to invert for the next model perturbation. This iterative process continues until the root mean square (RMS) error of the travel times falls below a predefined threshold.

## Features of pDSurfTomo

To achieve targeted parallel optimization of DSurfTomo, we first conduct a systematic profiling and modular decomposition of the original source code. Based on the code logic and computational characteristics, we partition the inversion workflow into three primary modules: sensitivity kernel module, traveltime module, and least-squares solver module. Non-computational routine, such as parameter input/output (I/O), data storage, and workflow control, are categorized into an auxiliary module. Specifically, the sensitivity kernel module calculates the sensitivity of the phase (or group) velocity with respect to the 3D S-wave velocity model to construct matrix $\boldsymbol{G}$; The traveltime module executes surface wave ray tracing and theoretical traveltime computation to construct the residual vector $\Delta\boldsymbol{d}$; The least-squares solver module is dedicated to solving the least-squares problem (Eq. 6). The execution time for each module is evaluated using the following benchmarking protocol: each module is executed ten times, and the final execution time is determined by averaging the results after discarding the maximum and minimum values. The hardware configuration of the experimental platform consists of two AMD EPYC 7643 48-Core Processors (totaling 96 physical cores and 192 logical processors) and an NVIDIA A100 40GB PCIe GPU. To maintain experimental rigor, all subsequent performance evaluations of the parallel optimizations are conducted using this identical hardware setup and timing standard.

Based on approximately 360,000 observed dispersion measurements acquired from a dense seismic array in North China (detailed in the Section of Application of pDSurfTomo to a dense array deployed in North China), the computation time for each module of DSurfTomo under a 64-thread parallel configuration is illustrated in Figure 1. The profiling results reveal that the three modules associated with the construction and solution of the linear system $\Delta \boldsymbol{d} = \boldsymbol{G}\Delta \boldsymbol{m}$ account for 99% of the total computation time, constituting the primary bottleneck that limits inversion efficiency. Therefore, we implement targeted computational optimizations for each of these modules. The architectural evolution of the algorithm, contrasting the original serial workflow with our proposed hybrid framework, is illustrated in Figure 2. Comprehensive implementation details for these three optimized modules are described below.

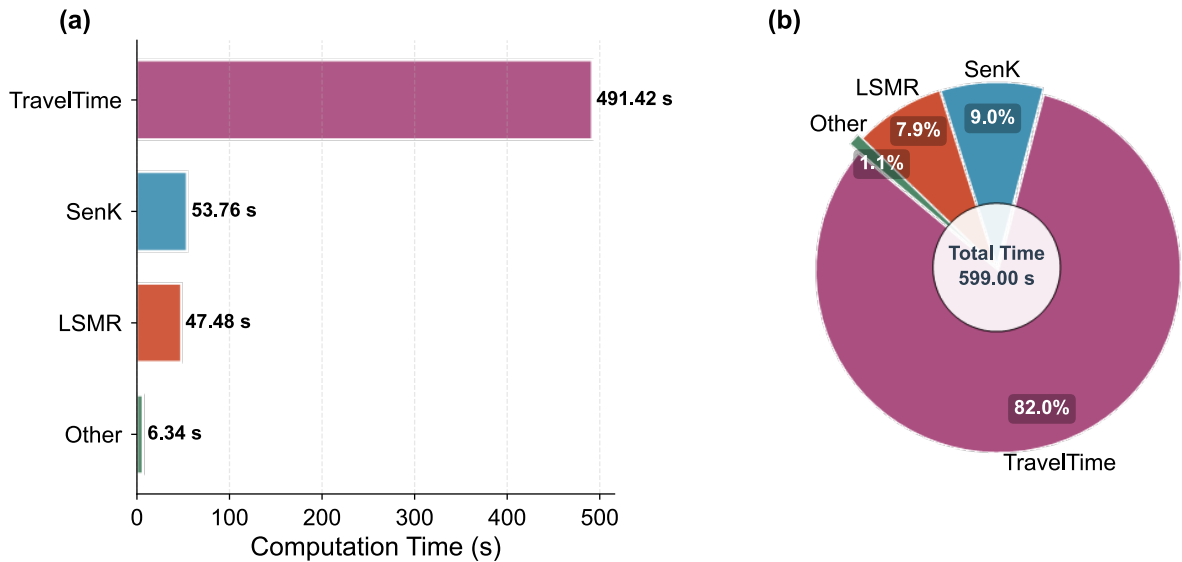

**Figure 1.** Computation time for each module of DSurfTomo. TravelTime, SenK, and LSMR represent the traveltime, sensitivity kernel, and least-squares solver modules, respectively; "Other"

denotes auxiliary modules. (a) Computation time for each module. (b) Percentage of total computation time for each module. The test is based on approximately 360,000 observed dispersion measurements acquired from a dense seismic array deployed in North China (detailed in the Section of Application of pDSurfTomo to a dense array deployed in North China).

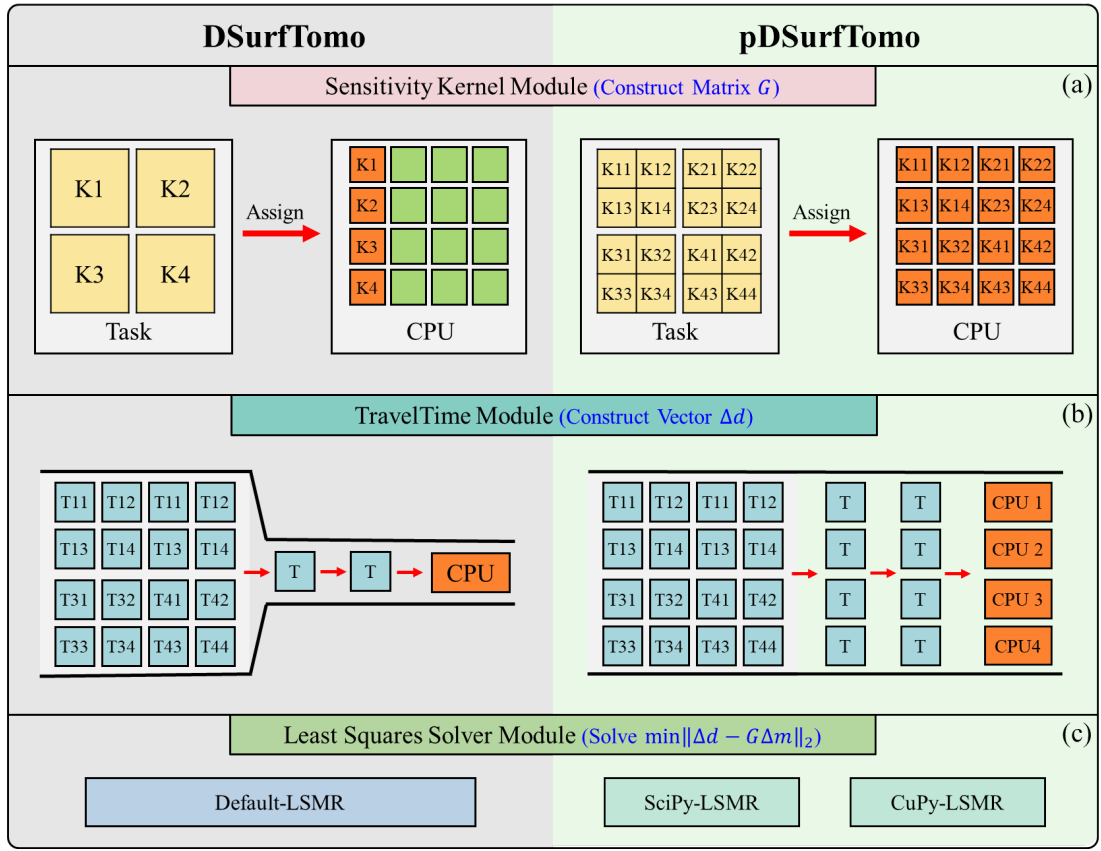

**Figure 2.** Architectural comparison between the original DSurfTomo and the proposed pDSurfTomo. (a) Computational strategies for the sensitivity kernel module. Blocks labeled "K" denote individual sensitivity kernel subtasks, where orange and green blocks denote active and idle CPU states, respectively. (b) Computational strategies for the traveltime module. Blocks labeled "T" represent independent traveltime calculation subtasks. (c) Comparison of the least-

squares solvers.

## Parallelization of Sensitivity Kernel Computation

The construction of the sensitivity kernel matrix $G$ relies on evaluating the partial derivatives of the phase (or group) velocity with respect to the local model parameters (i.e., P-wave velocity $\alpha$, S-wave velocity $\beta$, and density $\rho$) at specific depths. To evaluate these partial derivatives, DSurfTomo uses a perturbation-based central finite-difference scheme. For instance, considering the S-wave velocity $\beta_k(z_j)$ at depth $z_j$ beneath horizontal grid point $k$, the calculation of its sensitivity kernel $\partial C_{kj}(\omega)/\partial \beta_k(z_j)$ is given by:

$$\frac{\partial C_{kj}(\omega)}{\partial \beta_k(z_j)} = \frac{C_{kj2}(\omega) - C_{kj1}(\omega)}{2\Delta\beta_{kj}}, \qquad (7)$$

where $C_{kj1}(\omega)$ and $C_{kj2}(\omega)$ respectively donate the theoretical phase (or group) velocities obtained via forward modeling after applying negative and positive perturbations $\Delta\beta_{kj}$ to the S-wave velocity at depth $z_j$ beneath horizontal grid point $k$. Their specific forms are expressed as:

$$\begin{aligned} C_{kj1}(\omega) &= F(\alpha_k, \beta_k - \Delta\beta_{kj}, \rho_k, \omega) \\ C_{kj2}(\omega) &= F(\alpha_k, \beta_k + \Delta\beta_{kj}, \rho_k, \omega) \end{aligned}. \qquad (8)$$

where $\alpha_k$, $\beta_k$ and $\rho_k$ respectively denote the 1-D depth profiles of P-wave velocity, S-wave velocity, and density beneath horizontal grid point $k$; $F$ represents the surface wave dispersion forward modeling operator (Dunkin, 1965; Herrmann, 2013), which calculates the surface wave phase (or group) velocity at frequency $\omega$ for horizontal grid point $k$ based on the given 1-D layered model ($\alpha_k, \beta_k, \rho_k$). Similarly, the sensitivities of the phase (or group) velocity with respect to the P-wave velocity $\alpha$ and density $\rho$ at different depths can be computed sequentially. Ultimately, the full sensitivity matrix $\boldsymbol{G}$ is constructed by aggregating the sensitivities from all grids.

From a computational perspective, the perturbation-based sensitivity kernel evaluation in DSurfTomo is highly amenable to parallelization. Because the partial derivative computations at distinct spatial grid points and depth levels are mutually independent, they exhibit no inter-task data dependencies or algorithmic coupling. Consequently, substantial parallel acceleration can be realized through optimized task partitioning and efficient thread scheduling strategies. However, as shown in Figure 2a, the parallel design of the existing sensitivity kernel module in DSurfTomo exhibits certain deficiencies. Its low degree of parallelization fails to fully exploit the parallel computing capabilities of multicore CPUs. Specifically, the module adopts a simple coarse-grained task partitioning strategy generating fewer parallel tasks than the number of available CPU cores. Consequently, most CPU cores remain idle due to a lack of allocated tasks, while a few become overloaded, creating performance bottlenecks that significantly degrade the parallel efficiency of the sensitivity kernel computation. To address this issue, we build upon existing parallel architecture of DSurfTomo to implement a more refined parallel design for the sensitivity kernel module, and develop the improved SenK-Parallel module. First, we further decouple the sensitivity kernel module into two independent submodules for dispersion forward modeling and sensitivity kernel calculation, respectively. We execute the dispersion forward modeling tasks in parallel, and only after all dispersion curves are computed, we then collectively calculate the sensitivity kernels.

This approach ensures that memory associated with forward modeling is accessed intensively within a continuous timeframe. This strategy not only significantly improves CPU cache hit rates and mitigates performance degradation from cache misses, but also avoids frequent inter-process context switching caused by the interleaved execution of different tasks. Next, we adopt a fine-grained task partitioning strategy tailored to the computational characteristics of the sensitivity kernel. Specifically, using the 3D grid points (defined by horizontal grid point $k$ and depth $z_j$) as the fundamental units, we decompose the dispersion forward modeling into numerous independent subtasks, each dedicated to a parameter perturbation at a single grid point. Since calculating the sensitivity of model parameter for each grid point requires two forward modeling runs, computing the sensitivities of phase velocity to model parameters ($\alpha$, $\beta$, and $\rho$) for a 3-D model with a total of M grid points requires 6M independent subtasks. This task volume, which significantly outnumbers CPU cores, allows the sensitivity kernel module to dynamically allocate workloads in real time based on the operational status of the CPU cores. The dynamic scheduling strategy guarantees that all CPU cores remain fully saturated throughout the entire computation, thereby maximizing resource utilization and overall parallel efficiency. Furthermore, to accommodate the column-major memory layout of the Fortran programming language, we also optimize the data storage and access pattern. By enforcing a contiguous memory access pattern, we leverage spatial locality to substantially improve CPU cache hit rates. This approach effectively mitigates the cache misses and memory bandwidth bottlenecks conventionally induced by strided memory access.

Vectorization is also commonly used for enhancing computational efficiency. Essentially, vectorization eliminates explicit loops by transforming serial scalar operations into instruction-level parallel vector operations. To further reduce the computation time of sensitivity kernel, we adopt disba, a computationally efficient open-source Python library for modeling of surface wave dispersion (Luu, 2024), as our dispersion forward modeling operator. Building upon this foundation, we leverage the generalized universal function vectorization technique in Numba (Lam et al., 2015) to optimize the dispersion forward modeling kernel in disba. This optimization transforms the original scalar computing paradigm into a parallel vector computing paradigm, which facilitates highly efficient batch computation of surface wave dispersion and enables the development of the SenK-Disba module.

## Parallelization in TravelTime Computation

The construction of the traveltime residual vector $\Delta \boldsymbol{d}$ requires accurate theoretical surface wave traveltimes. DSurfTomo achieves this by using the FMM (Rawlinson and Sambridge, 2003), which effectively accounts for ray path bending effects at different periods. However, it should be noted that this improved accuracy comes at a higher computational cost. As shown by the performance profiling results in Figure 1, the traveltime module accounts for up to 80% of the total computation time, making it the primary bottleneck restricting overall inversion efficiency. The reason for this high computation time, as illustrated in Figure 2b, is that the FMM algorithm uses a purely serial architecture. It lacks support for multicore CPUs parallel environments, rendering it unable to leverage modern multicore processing to reduce computation time. Consequently, implementing a parallelized FMM is essential for accelerating the overall inversion process.

In terms of computational characteristics, FMM possesses significant inherent parallel potential. First, the 2D phase velocity maps for different periods are independent. The ray tracing for each period relies exclusively on its corresponding velocity field parameters, indicating there is no cross-period data coupling. Second, computing traveltimes for multiple ray paths within a single period requires only the 2D phase velocity map of that specific period. Apart from the 2D phase velocity map, the traveltime calculations for individual paths are completely decoupled and can be executed independently.

To implement parallel traveltime computation, selecting an appropriate parallel framework is essential. In scientific computing, the mainstream parallel programming models are OpenMP (Open Multi-Processing; Dagum and Menon, 1998) and MPI (Message Passing Interface; Walker and Dongarra, 1996). Based on shared-memory and distributed-memory architectures respectively, these frameworks cover a diverse range of parallel computing requirements, scaling from single-node multicore systems to large-scale computational clusters. Considering that the sensitivity kernel module in DSurfTomo is implemented using OpenMP, we adopt the same parallel framework for the parallel calculation of traveltimes to maintain architectural consistency. This choice effectively avoids potential compatibility conflicts associated with mixed programming models.

After the parallel computing architecture is established, we implement a targeted parallel scheme for the traveltime module based on the specific computational characteristics of surface waves. First, we refactor the original serial FMM to enable parallel execution. The original algorithm exhibits strong data dependencies, leading to data races and memory conflicts in multithreaded, shared-memory environments, which renders it unsuitable for direct parallelization. To address this, we decouple the internal data dependencies and optimize the data storage structures. This refactoring ensures thread safety and allows the algorithm to be executed independently across multiple threads. Then, we design a fine-grained task partitioning strategy to achieve efficient scheduling and load balancing. Specifically, using the surface wave period and ray path as dual partitioning criteria, we decompose the overall traveltime computation into numerous independent subtasks. Each subtask is dedicated to calculating the theoretical travel time for the $i$-th path at frequency $\omega$. For thread allocation, we use OpenMP compiler directives to dynamically assign thread resources. This approach allows each thread to independently execute one or more sub-tasks, effectively mitigating load imbalances caused by overloaded single threads or idle CPU cores. Furthermore, to guarantee data consistency in the parallel environment, we design an efficient thread synchronization mechanism. Each thread executes its tasks utilizing thread-local memory, and the final computational results are aggregated using a global barrier synchronization. This approach significantly minimizes the performance overhead typically caused by frequent synchronization. Through these designs, we achieve highly efficient parallelization of the traveltime module with the OpenMP programming model (Fig. 2b). This implementation not only fully leverages the parallel computing capabilities of multicore CPUs but also ensures the reliability of the ray tracing and traveltime computation results.

## Solver for the Large-scale Linear System

After constructing the sensitivity kernel matrix $\boldsymbol{G}$ and traveltime residual vector $\Delta\boldsymbol{d}$, the inverse problem is reduced to solving a large-scale sparse overdetermined linear system $\Delta\boldsymbol{d} = \boldsymbol{G}\Delta\boldsymbol{m}$. For such problems, the mainstream numerical solvers currently include the least squares with QR-factorization (LSQR) (Paige and Saunders, 1982) and the least squares with minimal residual (LSMR) (Fong and Saunders, 2011). LSMR and LSQR are numerically equivalent to applying the minimal residual method (Paige and Saunders, 1975) and the conjugate gradient method (Hestenes and Stiefel, 1952) to the normal equations, respectively. Owing to the superior convergence properties of the minimal residual approach, LSMR guarantees the simultaneous monotonic decrease of both the residual norm $\|\boldsymbol{r}\|$ and the normal equation residual norm $\|\boldsymbol{G}^T\boldsymbol{r}\|$ (here $\boldsymbol{r} = \Delta\boldsymbol{d} - \boldsymbol{G}\Delta\boldsymbol{m}$). In contrast, LSQR only guarantees the monotonic decrease of $\|\boldsymbol{r}\|$. This dual monotonic convergence property enables LSMR to achieve more robust convergence behavior and enhanced numerical stability when solving the ill-posed least-squares problems inherent to surface wave tomography. Consequently, DSurfTomo adopts LSMR as its linear solver.

It is worth mentioning that mainstream scientific computing libraries like SciPy (Virtanen et al., 2020) and CuPy (Okuta et al., 2017) have both implemented mature LSMR solvers (hereafter denoted as SciPy-LSMR and CuPy-LSMR, respectively). Both implementations significantly outperform the native DSurfTomo LSMR solver (DSurfTomo-LSMR) in computational efficiency. The LSMR solver within SciPy is built upon highly optimized underlying code. By interfacing with BLAS and LAPACK libraries, it achieves efficient instruction-level parallelism and memory management on CPU platforms. Meanwhile, CuPy, a scientific computing library tailored for NVIDIA GPUs, features an LSMR solver that offers complete compatibility with the SciPy API. CuPy exploits the massively parallel processing capabilities of CUDA cores by offloading the most computationally demanding operations, such as sparse matrix-vector multiplications and vector norm calculations, to the GPU for hardware acceleration.

Motivated by these capabilities, we propose a CPU-GPU dual-path acceleration strategy. This approach not only further reduces the inversion time of DSurfTomo but also achieves flexible adaptation of computational power across diverse hardware environments. On the one hand, we construct a highly efficient CPU computational path based on SciPy-LSMR. By replacing the native solver, we exploit the mature algorithmic implementation and underlying hardware optimizations of SciPy-LSMR to substantially accelerate the solution of the least-squares problem. On the other hand, we further develop a massively parallel GPU acceleration capability using CuPy-LSMR, which harnesses the immense parallel processing power of thousands of GPU cores to overcome computational bottlenecks and reduce the time required to solve large-scale sparse linear least-squares problems.

## User-friendliness through cross-platform GUI interface

In practical applications, large-scale surface wave tomography is typically performed on remote dedicated servers to leverage their high-performance computing capabilities. However, since remote servers predominantly operate via command line interfaces and lack GUI, geophysicists are forced to configure parameters and visualize inversion results through terminal windows. This operational mode not only increases complexity and the likelihood of human error, but also imposes a steep learning curve on users.

To address these limitations, we develop cross-platform software for surface wave tomography based on Python and the PyQt5, which operates robustly on both Windows and Linux operating systems. The software features a remote dedicated server connection module. Through this module, users can configure parameters, execute tasks, and visualize results on a remote server entirely through the local GUI, bypassing the need to interact directly with the server terminal. This convenience effectively mitigates the learning curve and operational difficulty while ensuring the efficient of inversion tasks. The functionalities of this software are systematically implemented through the following five specialized interfaces.

(1) Main Inversion Interface: As shown in Figure 3, this interface enables the graphical execution of inversion tasks through the integration of seven functional panels.

Specifically, the model parameter configuration panel (Fig. 3a) defines the grid discretization and initial velocity model, with the results shown in Figure 3e and Figure 3f, respectively. The parallel parameter configuration panel (Fig. 3b) manages the parallel settings for the sensitivity kernel, traveltime, and least-squares solver modules, while the inversion parameter configuration panel (Fig. 3c) is dedicated specifically to the inversion settings. To ensure robust execution, the log monitoring panel (Fig. 3d) tracks runtime logs and standard outputs in real time, with errors highlighted in red for rapid troubleshooting of parameter or computational faults. Finally, the task control panel (Fig. 3g) manages the execution and termination of the inversion tasks.

(2) Observation System Interface: As illustrated in Figure S1, this interface serves as a supplementary tool for the validation of the observation system. By clearly exhibiting the spatial distribution of stations, the coverage of surface wave ray paths, and the grid discretization, the interface enables users to rapidly assess the adequacy of the observation system. Furthermore, this interface features interactive grid adjustment capabilities, allowing users to dynamically modify the grid resolution and spatial extent to achieve a precise geometric match between the observation system and the inversion grid, which ensures the reliability of the inversion results.

(3) MultiSlice Visualization Interface: As shown in Figure S2, this interface is engineered for real-time visualization and image export of inversion results. The left panel features an auto-updating file directory that continuously shows the output files generated after each iteration. The primary rendering area facilitates the interactive visualization of the 3-D velocity model via horizontal slices at multiple depths. When a specific file is selected from the directory, this area dynamically updates to display the slices for the corresponding iteration. Through the bottom control panel, users can flexibly adjust plotting parameters, such as colormaps and color scale limits (vmin and vmax), to precisely delineate velocity anomalies within the model. Furthermore, this interface integrates a multi-format image export function to generate publication-ready figures.

(4) Orthogonal Slice Visualization Interface: As illustrated in Figure S3, this interface enables the dynamic inspection of the 3-D velocity model along three orthogonal axes (one horizontal slice and two mutually perpendicular vertical cross-sections). Users can interactively adjust the spatial positions of these slices using three sets of sliders at the bottom, thereby facilitating an intuitive examination of the 3-D subsurface heterogeneity.

(5) SSH Interface: As depicted in Figure S4, this interface is specifically tailored for high-performance computing environments, establishing a secure communication channel between the local client and the remote server. Upon successful connection, users can seamlessly manage and execute remote inversion tasks entirely through local graphical interactions, bypassing the need to interact directly with the server terminal.

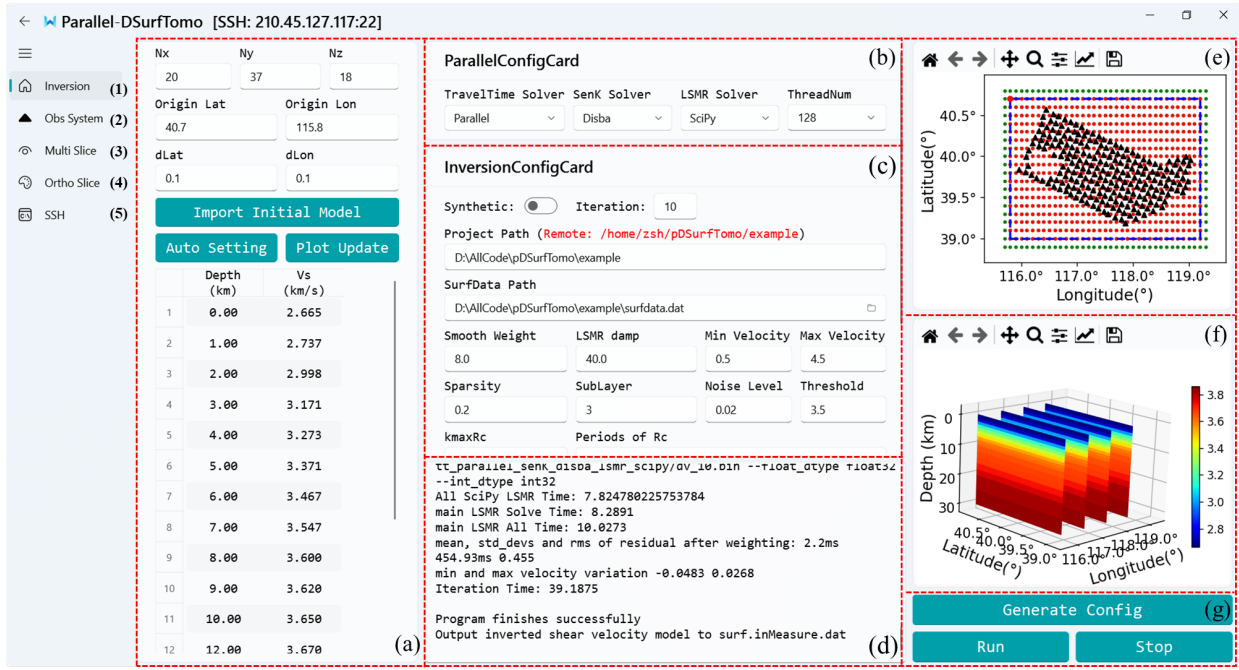

**Figure 3.** Main Inversion Interface. (a) Model parameter configuration panel. (b) Parallel parameter configuration panel. (c) Inversion parameter configuration panel. (d) Log monitoring panel. (e) Observation system visualization panel, which shows the spatial distribution of the seismic array and the horizontal grid points of the velocity model. (f) Model visualization panel, illustrating the initial 3D S-wave velocity model. (g) Task control panel, managing the execution and termination of the inversion tasks. Sidebar buttons (1-5) correspond to the Main Inversion Interface, Observation System Interface, MultiSlice Visualization Interface, Orthogonal Slice Visualization Interface, and SSH Interface, respectively.

# Application of pDSurfTomo to a dense array deployed in North China

To validate the efficacy of pDSurfTomo, we systematically evaluate its computational performance and inversion reliability. This validation utilizes observed dispersion data derived from ambient noise recordings acquired by a dense seismic array deployed in North China. The array comprised 229 SmartSolo IGU-16HR 3C nodal seismometers, which continuously recorded ambient noise signals for a one-month period from November to December 2023. The spatial distribution of these stations, along with the primary topographic features of the study region, is depicted in Figure 4a. The seismic array spans an area of approximately 90 km × 230 km with an average inter-station spacing of 10 km, which effectively encompasses diverse tectonic units within the region. This dense coverage provides a robust observational foundation for extracting high-quality surface wave dispersion data and conducting subsequent tomographic imaging. Utilizing these recordings, we manually extracted Rayleigh wave phase velocity dispersion data in the periods of 1.0 to 8.0 s using the ambient noise cross-correlation method (Yao et al., 2006; Deng et al., 2022; Ni et al., 2023). The extracted dispersion curves and their distribution characteristics are illustrated in Figure 4c. Horizontally, the study region is discretized into a 20 × 37 grid with a uniform interval of 0.015°; vertically, the model is parameterized by 18 nodes with non-uniform spacing. This depth configuration and the initial 1-D S-wave velocity profile are presented in Figure 4b.

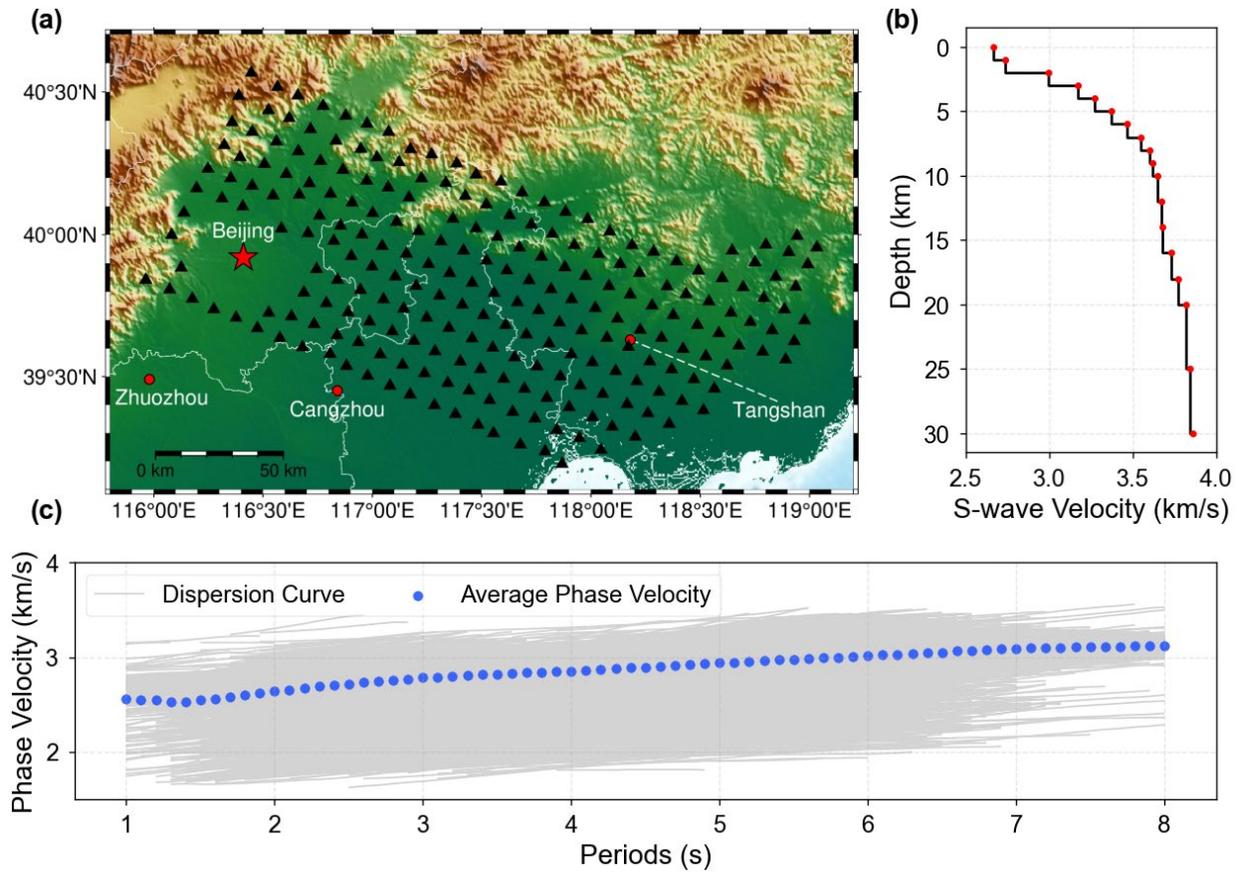

**Figure 4.** Overview of the dispersion data used for validating pDSurfTomo. (a) Distribution of stations and the primary topographic features of the study region. (b) Depth profile of the initial 1-D S-wave velocity model used for inversion. (c) Rayleigh wave phase velocity dispersion data within the study region.

## Performance Evaluation of pDSurfTomo

In this section, we evaluate the computational performance of the sensitivity kernel module, traveltime module, and the least-squares solvers, using the observed dispersion data from the dense array deployed in North China. The benchmarking protocol is defined as follows: the execution time for each module is recorded over ten iterations, and the final execution time is calculated as the average excluding the maximum and minimum values. To provide a quantitative benchmark of the algorithmic improvements, we define the acceleration factor (Acc-Factor) as the ratio of the execution time of a native DSurfTomo module to that of its optimized pDSurfTomo equivalent. For the sensitivity kernel and traveltime modules, the ratio of execution time is calculated under identical thread counts. For the least-squares solver, however, the original DSurfTomo implementation does not permit explicit thread allocation. Therefore, its performance baseline is established by comparing the execution time of the native DSurfTomo-LSMR with that of our optimized SciPy-LSMR and CuPy-LSMR solvers.

The computation time and acceleration factor of the three modules are presented in Figure 5. Figure 5a illustrates the computation time and acceleration factor of the sensitivity kernel module under different thread counts. This figure highlights the performance disparities between the native and our optimized module, empirically validating the efficacy of our proposed optimization strategies. For the native DSurfTomo sensitivity kernel module, computation time levels off and exhibits slight fluctuations beyond 64 threads, which clearly indicates the limitations of its simple coarse-grained task partitioning. In contrast, the SenK-Parallel module, which incorporates a refined parallel design, not only reduces the computation time for a given thread count, but also successfully overcomes the CPU scalability bottleneck inherent in the native module. Consequently, its computation time consistently decreases as the number of allocated CPU threads scales up. Furthermore, by replacing the native dispersion forward modeling operator with the vectorized implementation of disba, the computation time of the sensitivity kernel module is further reduced by nearly a factor of 2 to 3 relative to the SenK-Parallel module under identical thread counts. Figure 5b presents the computation time for the traveltime module. Since FMM in DSurfTomo is implemented with a purely serial architecture, its computation time is independent of the CPU thread count, as shown by the flat red line. In contrast, pDSurfTomo implements a highly efficient parallelization of FMM, which not only significantly reduces the computation time of the traveltime module but also achieves a quasilinear decrease in computation time with an increasing number of CPU threads. Figure 5c presents the computation time for the least-squares solver module. Benefiting from highly optimized underlying algorithms in SciPy, SciPy-LSMR achieves a significant improvement in solver efficiency within a CPU-only environment, reducing the computation time by 75% compared to the native LSMR solver. Meanwhile, CuPy-LSMR leverages GPU-based heterogeneous parallel acceleration, which reduces the computation time by nearly a factor of 10 compared to the native LSMR solver, significantly enhancing the efficiency of solving large-scale sparse least-squares problems.

Beyond individual module benchmarks, we further conduct a systematic analysis of the per-iteration overall performance of both packages (Fig. 6). The computation time for pDSurfTomo is obtained under a CPU-based configuration, utilizing the SenK-Disba, TravelTime-Parallel, and SciPy-LSMR modules. Figure 6a shows the absolute computation time, and Figure 6b shows the corresponding acceleration factor. The trends in per-iteration runtime reveal that the performance of DSurfTomo is constrained by inherent limitations in the parallel design of its sensitivity kernel module, exhibiting a pronounced CPU scalability bottleneck. In contrast, pDSurfTomo successfully overcomes this bottleneck, allowing the iteration time to continuously decrease as the number of CPU threads increases. Furthermore, pDSurfTomo demonstrates a highly significant acceleration compared to DSurfTomo, achieving an acceleration factor of 10 or greater. Crucially, this factor continues to scale as the number of CPU threads further increases.

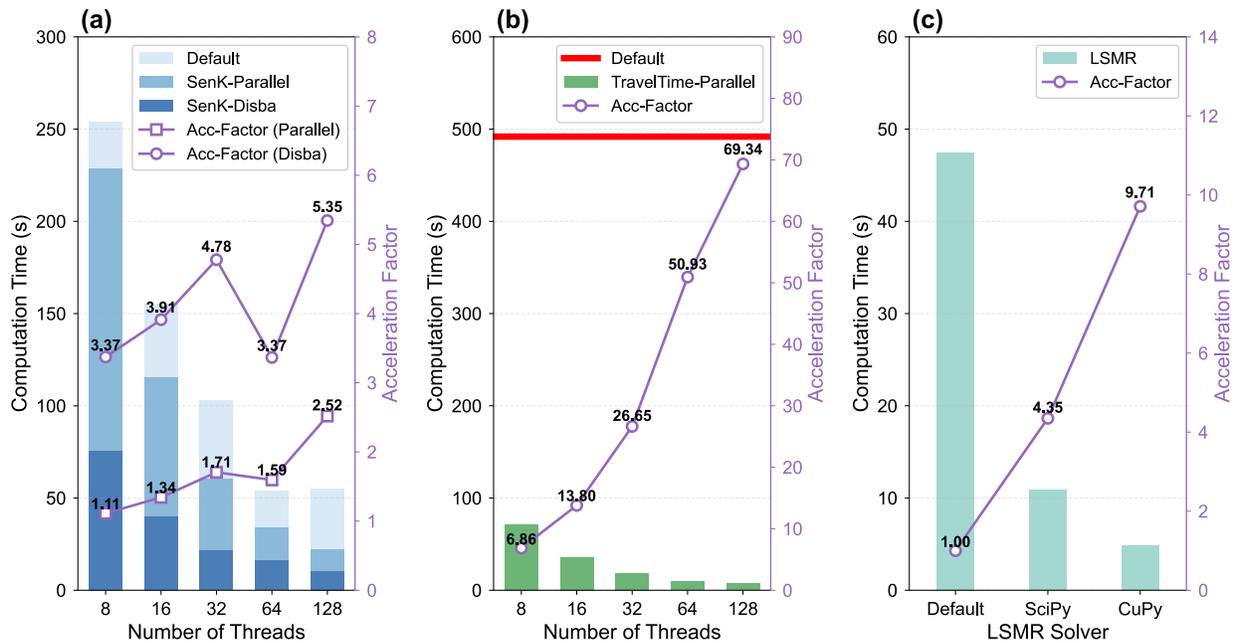

**Figure 5.** Computation time comparison of individual modules between the pDSurfTomo and DSurfTomo. "Default" denotes the native module of DSurfTomo. (a) Computation time and acceleration factor of the sensitivity kernel module under different thread counts. (b) Computation time and acceleration factor of the traveltime module under different thread counts. (c)

Computational times and acceleration factors of the three least-squares solvers. Note that none of the evaluated solvers (native DSurfTomo-LSMR, SciPy-LSMR, and CuPy-LSMR) permit explicit user-level thread allocation; instead, thread concurrency is managed internally by their respective backend mathematical libraries.

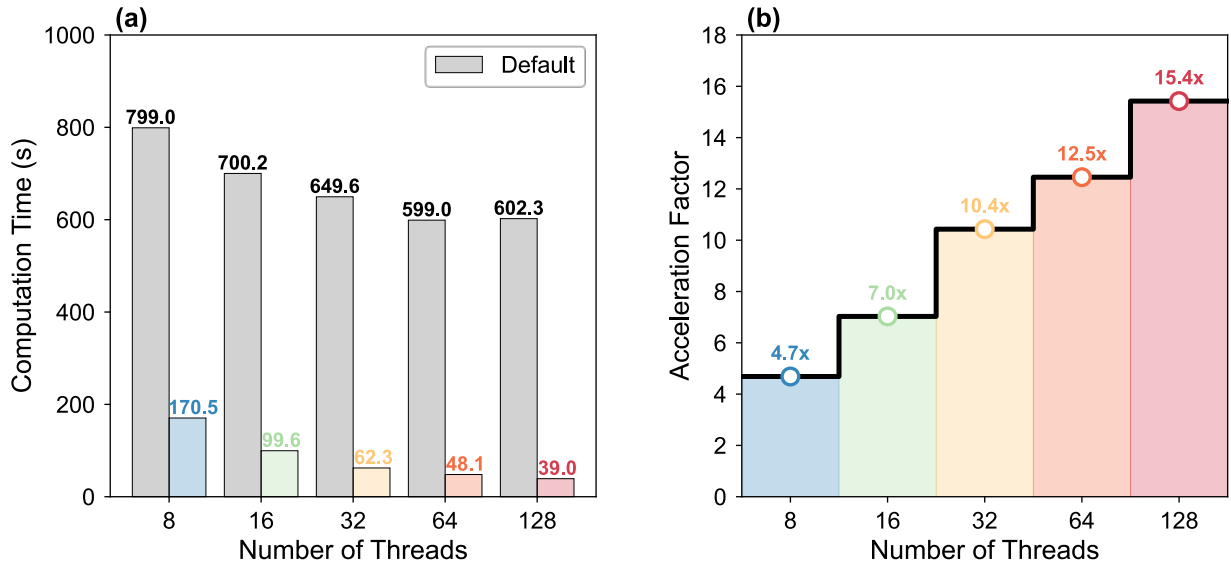

**Figure 6.** Comparison of the computation time per iteration between the pDSurfTomo and DSurfTomo. (a) Comparison times of both algorithms under different thread counts. Gray bars represent the Comparison time of DSurfTomo, while colored bars denote that of pDSurfTomo. (b) The corresponding acceleration factor achieved by pDSurfTomo under different thread counts.

## Reliability Evaluation of pDsurfTomo

Beyond evaluating computational performance, it is equally important to validate the inversion reliability of pDSurfTomo. Thus, we systematically compare the final 3D velocity models derived by both packages using the observed dispersion data from the dense seismic array deployed in North China. To ensure a fair comparison, both inversion workflows are initialized with an identical 1-D layer-stratified velocity model shown in Figure 4b.

Figure 7 illustrates the horizontal S-wave velocity slices of the final inverted models from both packages at depths of 6 km, 12 km, and 18 km. The comparison suggests that the inversion results are highly consistent. The largest discrepancy is on the order of $10^{-4}$, which is primarily attributed to different dispersion forward modeling operators and LSMR solvers.

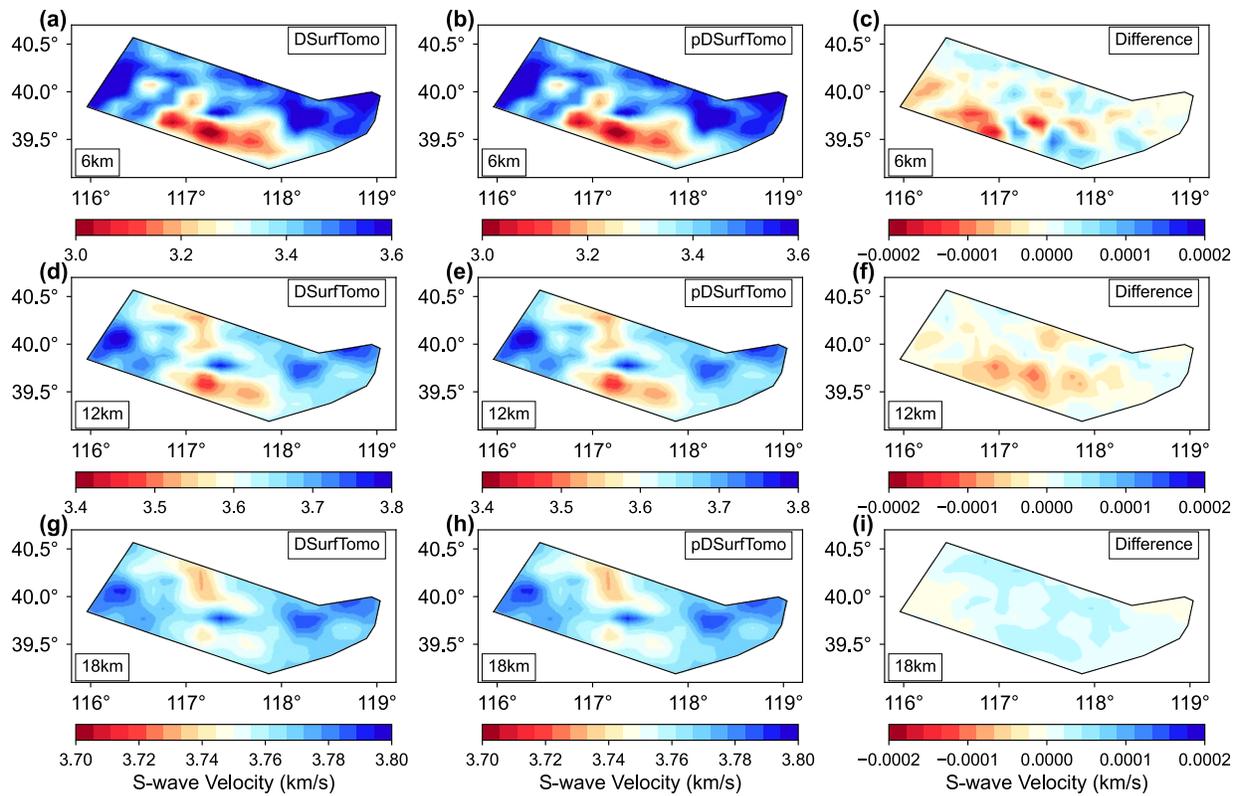

**Figure. 7.** Comparison of the inversion results by DSurfTomo and pDSurfTomo. (a)–(c) Horizontal slices of the inverted S-wave velocity model at the depth of 6 km by DSurfTomo, pDSurfTomo, and the difference between the two models, respectively. (d)–(f) are similar to (a)-(c) but for the depth of 12 km, and (g)–(i) are similar to (a)-(c) but for the depth of 18 km.

# Conclusion

Despite its widespread adoption in the community, the original DSurfTomo package has been constrained by inherent computational bottlenecks, limiting its applicability to large-scale, high-resolution surface wave tomography. To resolve these limitations, we develop pDSurfTomo, a highly optimized, next-generation package equipped with a modern frontend-backend architecture. pDSurfTomo overcomes the CPU scalability limits of sensitivity kernel computations through refined parallel design and vectorization, while a fine-grained parallelization of the fast marching method significantly accelerates theoretical traveltime computations. Furthermore, by integrating modern and GPU-accelerated sparse matrix solvers, the large-scale linear least-squares problems inherent to tomography are solved with satisfactory efficiency. Application to an observed dispersion dataset acquired by a dense seismic array deployed in North China confirms that pDSurfTomo achieves an order-of-magnitude reduction in computation time compared to the original software. Additionally, the software features a cross-platform graphical user interface with remote server connectivity, which allows users to manage high-performance computing tasks entirely from a local client. It is expected that pDSurfTomo will provide a robust, high-performance computational framework that benefits the community of surface wave tomography.

## Data and Resources

The pDSurfTomo software package, along with a subset of the data used in this study, is open-source and freely available at https://github.com/zshang825/pDSurfTomo. The original DSurfTomo package can be accessed at https://github.com/HongjianFang/DSurfTomo. The disba library used for surface wave dispersion forward modeling is available at https://github.com/keurfonluu/disba. Furthermore, we acknowledge the use of several open-source Python libraries that facilitated the development of this software. The graphical user interface (GUI) was constructed using PyQt-Fluent-Widgets (https://github.com/zhiyiYo/PyQt-Fluent-Widgets). The optimized least-squares solvers rely on SciPy (Virtanen et al., 2020) for CPU-based computations and CuPy (Okuta et al., 2017) for GPU acceleration. All websites and software repositories were last accessed in April 2026.

## Declaration of Competing Interests

The authors acknowledge that there are no conflicts of interest recorded.

## Acknowledgments

This study is financially supported by the National Key Research and Development Project of China (Grant No. 2022YFF0800701) and the Deep Earth Probe and Mineral Resources Exploration-National Science and Technology Major Project (Grant No. 2025ZD1007502).

# References

Boschi, L., & Ekström, G. (2002). New images of the Earth's upper mantle from measurements of surface wave phase velocity anomalies. Journal of Geophysical Research: Solid Earth, 107(B4), ESE-1.

Brocher, T. M. (2005). Empirical relations between elastic wavespeeds and density in the Earth's crust. Bulletin of the seismological Society of America, 95(6), 2081-2092.

Chen, G., Li, J., & Deng, B. (2026). Accurate Interpolation of Ambient Noise Empirical Green's Functions by Denoising Diffusion Probabilistic Model and Implicit Neural Representation. arXiv preprint arXiv:2601.05489.

Chen, K. X., Kuo‐Chen, H., Brown, D., Li, Q., Ye, Z., Liang, W. T., ... & Yao, H. (2016). Three‐dimensional ambient noise tomography across the Taiwan Strait: The structure of a magma‐poor rifted margin. Tectonics, 35(8), 1782-1792.

Dagum, L., & Menon, R. (1998). OpenMP: an industry standard API for shared-memory programming. IEEE computational science and engineering, 5(1), 46-55.

Deng, B., Li, J., Liu, J., Shen, C., Suwen, J., & Chen, Q. F. (2022). The extended range phase shift method for broadband surface wave dispersion measurement from ambient noise and its application in ore deposit characterization. Geophysics, 87(3), JM29-JM40.

Dunkin, J. W. (1965). Computation of modal solutions in layered, elastic media at high frequencies. Bulletin of the Seismological Society of America, 55(2), 335-358.

Fang, H., Yao, H., Zhang, H., Huang, Y. C., & van der Hilst, R. D. (2015). Direct inversion of surface wave dispersion for three-dimensional shallow crustal structure based on ray tracing: methodology and application. Geophysical Journal International, 201(3), 1251-1263.

Feng, M., & An, M. (2010). Lithospheric structure of the Chinese mainland determined from joint inversion of regional and teleseismic Rayleigh‐wave group velocities. Journal of Geophysical Research: Solid Earth, 115(B6).

Fong, D. C. L., & Saunders, M. (2011). LSMR: An iterative algorithm for sparse least-squares problems. SIAM Journal on Scientific Computing, 33(5), 2950-2971.

Herrmann, R. B. (2013). Computer programs in seismology: An evolving tool for instruction and research. Seismological Research Letters, 84(6), 1081-1088.

Hestenes, M. R., & Stiefel, E. (1952). Methods of conjugate gradients for solving linear systems. Journal of research of the National Bureau of Standards, 49(6), 409-436.

Hu, S., Yao, H., & Huang, H. (2020). Direct surface wave radial anisotropy tomography in the crust of the eastern Himalayan syntaxis. Journal of Geophysical Research: Solid Earth, 125(5), e2019JB018257.

Lam, S. K., Pitrou, A., & Seibert, S. (2015, November). Numba: A llvm-based python jit compiler. In Proceedings of the Second Workshop on the LLVM Compiler Infrastructure in HPC (pp. 1-6).

Li, J., Xu, J., Zhang, H., Yang, W., Tan, Y., Zhang, F., ... & Sun, J. (2023). High seismic velocity structures control moderate to strong induced earthquake behaviors by shale gas development. Communications Earth & Environment, 4(1), 188.

Li, L., Cai, C., Fang, Y., & Fang, H. (2022). Multiple surface wave tomography methods and their applications to the Tibetan Plateau. Reviews of Geophysics and Planetary Physics, 54(2), 174-196.

Li, W., He, R., Yuan, X., Schneider, F., Tilmann, F., Guo, Z., & Chen, Y. J. (2024). Correlated crustal and mantle melting documents proto-Tibetan Plateau growth. National Science Review, 11(9), nwae257.

Liu, C., Yao, H., Yang, H. Y., Shen, W., Fang, H., Hu, S., & Qiao, L. (2019). Direct inversion for three‐dimensional shear wave speed azimuthal anisotropy based on surface wave ray tracing: Methodology and application to Yunnan, southwest China. Journal of Geophysical Research: Solid Earth, 124(11), 11394-11413.

Luu, K. (2024). disba: Numba-accelerated computation of surface wave dispersion (v0.7.0). Zenodo. https://doi.org/10.5281/zenodo.14534395

Ni, H., Zhang, R., Li, J., Huang, X., Zheng, H., Hong, D., ... & Bao, Z. (2023). Application of horizontal-to-vertical spectral ratio on prospecting urbansite response and subsurface active faults in Mingguang, Anhui province. Chinese Journal of Geophysics, 66(11), 4552-4571.

Nimiya, H., Ikeda, T., & Tsuji, T. (2020). Three‐dimensional S wave velocity structure of central Japan estimated by surface‐wave tomography using ambient noise. Journal of Geophysical Research: Solid Earth, 125(4), e2019JB019043.

Nthaba, B., Ikeda, T., Nimiya, H., Tsuji, T., & Iio, Y. (2022). Ambient noise tomography for a high-resolution 3D S-wave velocity model of the Kinki Region, Southwestern Japan, using dense seismic array data. Earth, Planets and Space, 74(1), 96.

Okuta, R., Unno, Y., Nishino, D., Hido, S., & Loomis, C. (2017). CuPy: A NumPy-compatible library for NVIDIA GPU calculations. In Proceedings of Workshop on Machine Learning Systems (LearningSys) in the Thirty-first Annual Conference on Neural Information Processing Systems (NIPS).

Paige, C. C., & Saunders, M. A. (1975). Solution of sparse indefinite systems of linear equations. SIAM journal on numerical analysis, 12(4), 617-629.

Paige, C. C., & Saunders, M. A. (1982). LSQR: An algorithm for sparse linear equations and sparse least squares. ACM Transactions on Mathematical Software (TOMS), 8(1), 43-71.

Rawlinson, N., & Sambridge, M. (2003). Seismic traveltime tomography of the crust and lithosphere. Advances in geophysics, 46, 81-199.

Shapiro, N. M., & Ritzwoller, M. H. (2002). Monte-Carlo inversion for a global shear-velocity model of the crust and upper mantle. Geophysical Journal International, 151(1), 88-105.

Trampert, J., & Woodhouse, J. H. (1995). Global phase velocity maps of Love and Rayleigh waves between 40 and 150 seconds. Geophysical Journal International, 122(2), 675-690.

Virtanen, P., Gommers, R., Oliphant, T. E., Haberland, M., Reddy, T., Cournapeau, D., ... & Van Mulbregt, P. (2020). SciPy 1.0: fundamental algorithms for scientific computing in Python. Nature methods, 17(3), 261-272.

Walker, D. W., & Dongarra, J. J. (1996). MPI: a standard message passing interface. Supercomputer, 12, 56-68.

Woodhouse, J. H., & Dziewonski, A. M. (1984). Mapping the upper mantle: Three‐dimensional modeling of Earth structure by inversion of seismic waveforms. Journal of Geophysical Research: Solid Earth, 89(B7), 5953-5986.

Xiang, L., Huajian, Y., Yu, L., & Qiyuan, L. (2015). Effect of off-great-circle propagation on surface wave phase velocity tomography in western Sichuan. Acta Seismologica Sinica, 37(1), 15-28.

Yang, H. Y., & Hung, S. H. (2005). Validation of ray and wave theoretical travel times in heterogeneous random media. Geophysical research letters, 32(20).

Yao, H., van Der Hilst, R. D., & De Hoop, M. V. (2006). Surface-wave array tomography in SE Tibet from ambient seismic noise and two-station analysis—I. Phase velocity maps. Geophysical Journal International, 166(2), 732-744.

Yao, H., Beghein, C., & Van Der Hilst, R. D. (2008). Surface wave array tomography in SE Tibet from ambient seismic noise and two-station analysis-II. Crustal and upper-mantle structure. Geophysical Journal International, 173(1), 205-219.

Yao, H., van Der Hilst, R. D., & Montagner, J. P. (2010). Heterogeneity and anisotropy of the lithosphere of SE Tibet from surface wave array tomography. Journal of Geophysical Research: Solid Earth, 115(B12).

Yao, H., Luo, S, L., Li C., Hu S., & Fang H. (2022). Direct surface wave tomography for three dimensional structure based on surface wave traveltimes: Methodology review and applications. Reviews of Geophysics and Planetary Physics, 54(3), 231-251.

Zhai, B., Deng, J., Yu, H., Wang, X., Chen, H., & Wang, Q. (2025). Shallow crustal velocity structure beneath the Xiangshan and Yuhuashan volcanic basins in South China: Implications for the metallogenic setting of the volcanic-related uranium deposit. Ore Geology Reviews, 180, 106551.

Zhao, Y., Li, J., Xu, J., Yao, H., Zhu, G., Yang, H., ... & Lu, R. (2023). High-resolution velocity structure and seismogenic potential of strong earthquakes in the Bamei-Kangding segment of the Xianshuihe fault zone. Science China Earth Sciences, 66(9), 1960-1978.

Zhou, M., Tian, X., Wang, F., Wei, Y., & Xin, H. (2019). Shallow velocity structure of the Luoyang basin derived from dense array observations of urban ambient noise. Earthquake Science, 31(5-6), 252-261.

# Figures

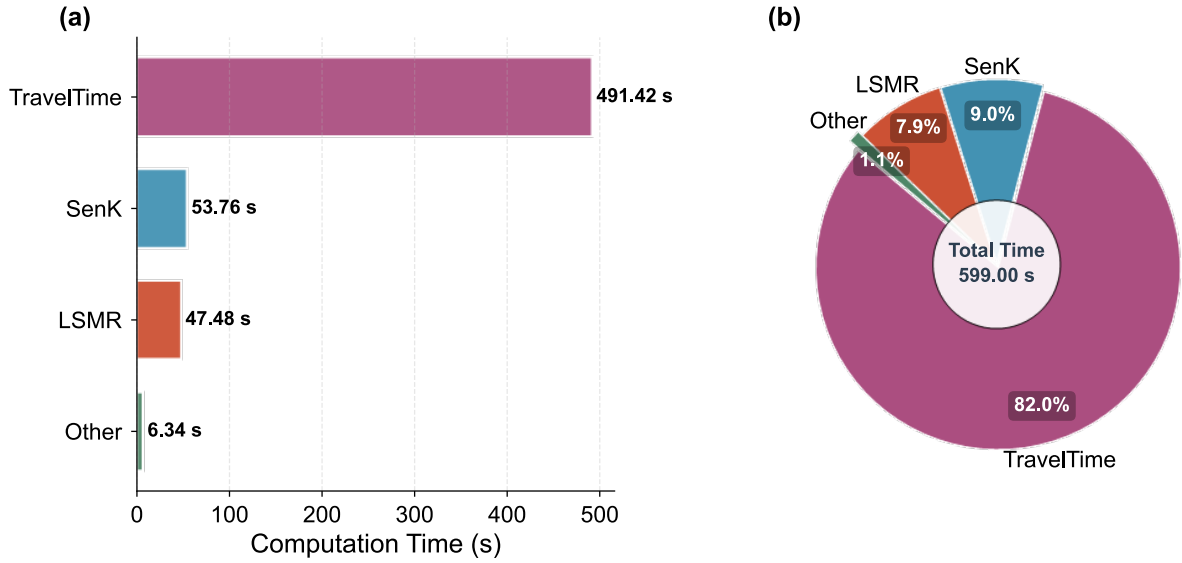

**Figure 1.** Computation time for each module of DSurfTomo. TravelTime, SenK, and LSMR represent the traveltime, sensitivity kernel, and least-squares solver modules, respectively; "Other" denotes auxiliary modules. (a) Computation time for each module. (b) Percentage of total computation time for each module. The test is based on approximately 360,000 observed dispersion measurements acquired from a dense seismic array deployed in North China (detailed in the Section of Application of pDSurfTomo to a dense array deployed in North China).

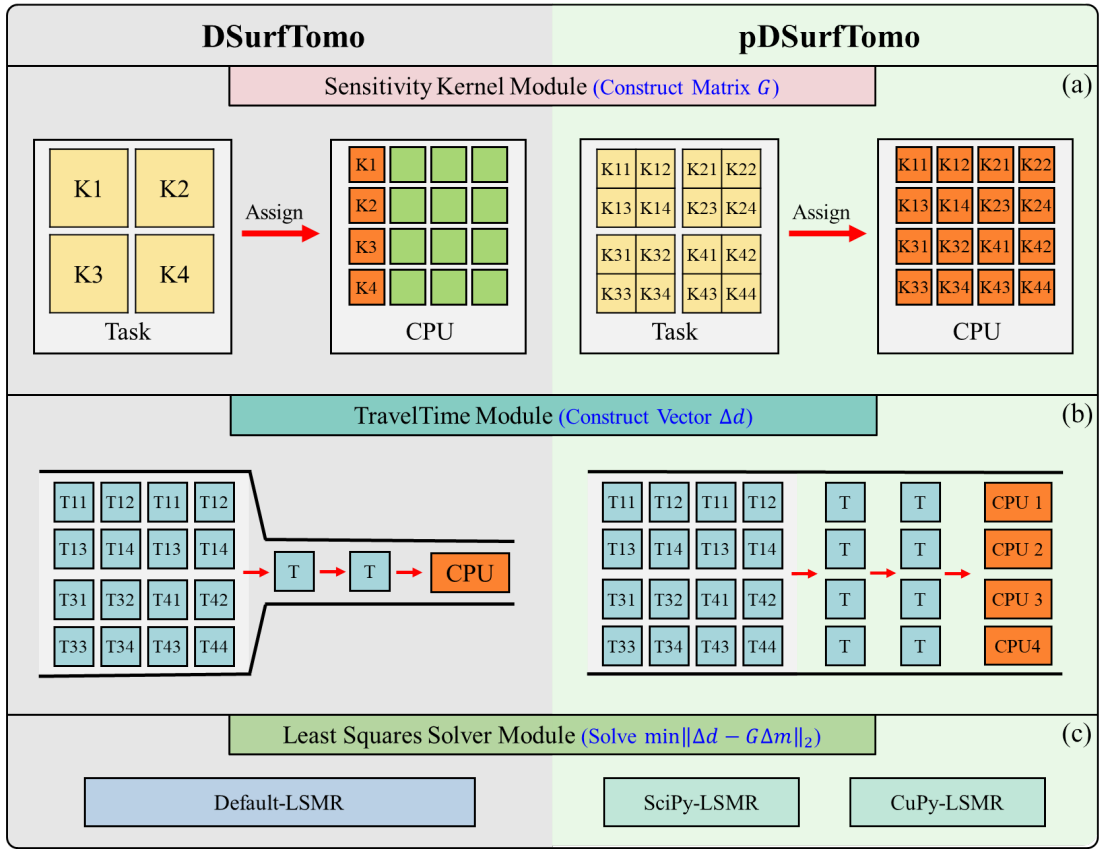

**Figure 2.** Architectural comparison between the original DSurfTomo and the proposed pDSurfTomo. (a) Computational strategies for the sensitivity kernel module. Blocks labeled "K" denote individual sensitivity kernel subtasks, where orange and green blocks denote active and idle CPU states, respectively. (b) Computational strategies for the traveltime module. Blocks labeled "T" represent independent traveltime calculation subtasks. (c) Comparison of the least-squares solvers.

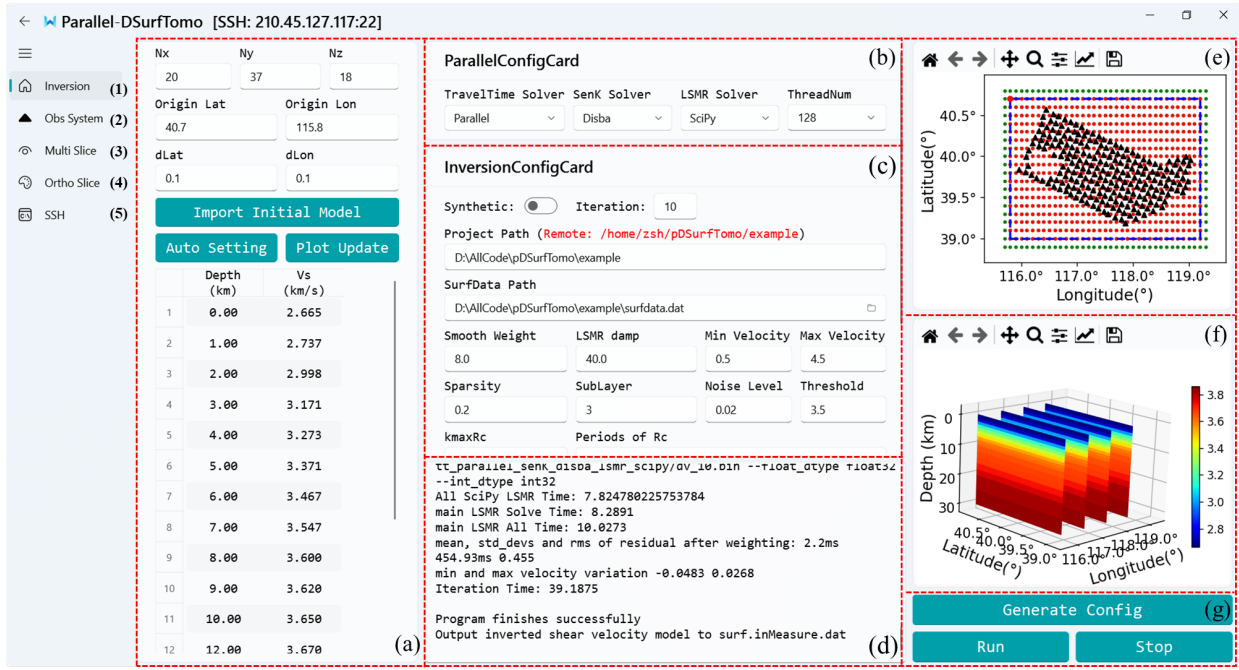

**Figure 3.** Main Inversion Interface. (a) Model parameter configuration panel. (b) Parallel parameter configuration panel. (c) Inversion parameter configuration panel. (d) Log monitoring panel. (e) Observation system visualization panel, which shows the spatial distribution of the seismic array and the horizontal grid points of the velocity model. (f) Model visualization panel, illustrating the initial 3D S-wave velocity model. (g) Task control panel, managing the execution and termination of the inversion tasks. Sidebar buttons (1-5) correspond to the Main Inversion Interface, Observation System Interface, MultiSlice Visualization Interface, Orthogonal Slice Visualization Interface, and SSH Interface, respectively.

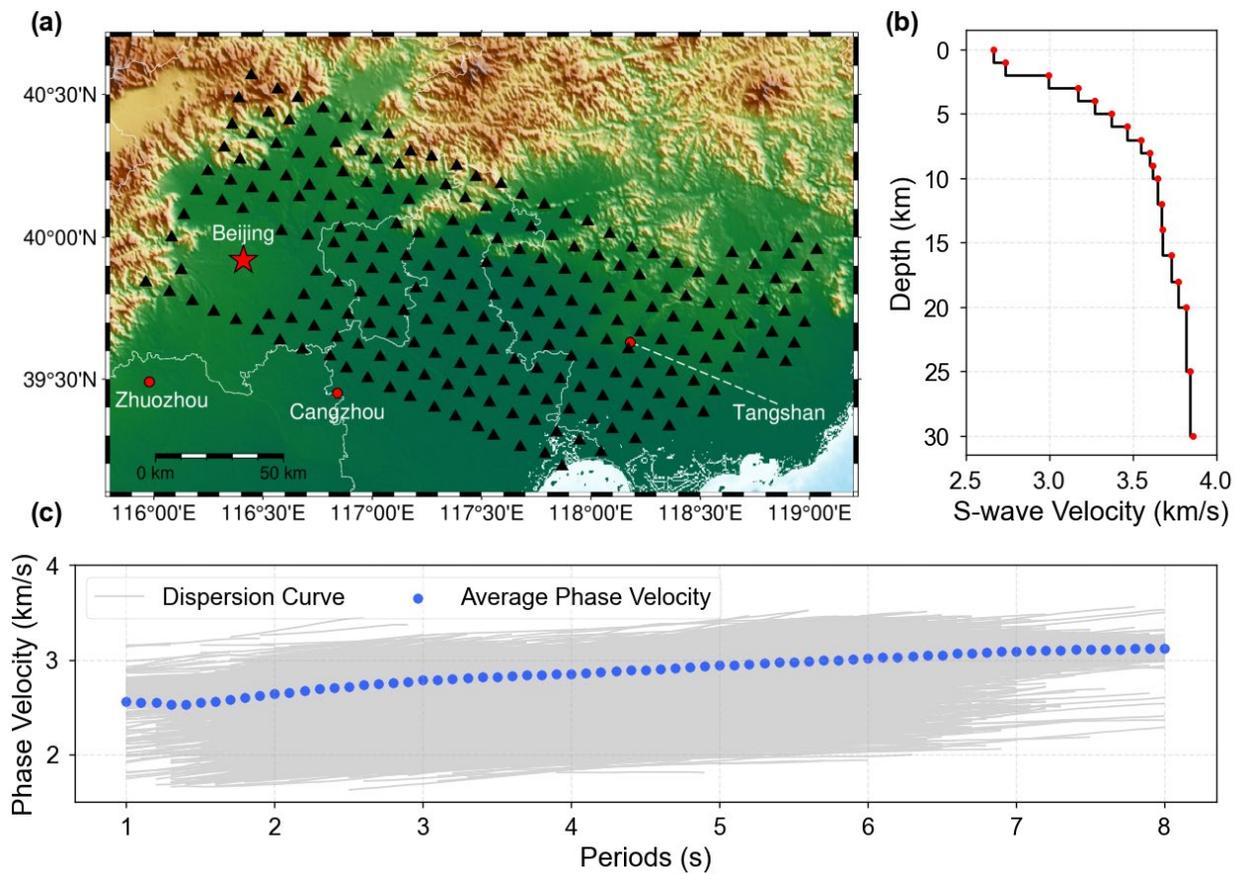

**Figure 4.** Overview of the dispersion data used for validating pDSurfTomo. (a) Distribution of stations and the primary topographic features of the study region. (b) Depth profile of the initial 1-D S-wave velocity model used for inversion. (c) Rayleigh wave phase velocity dispersion data within the study region.

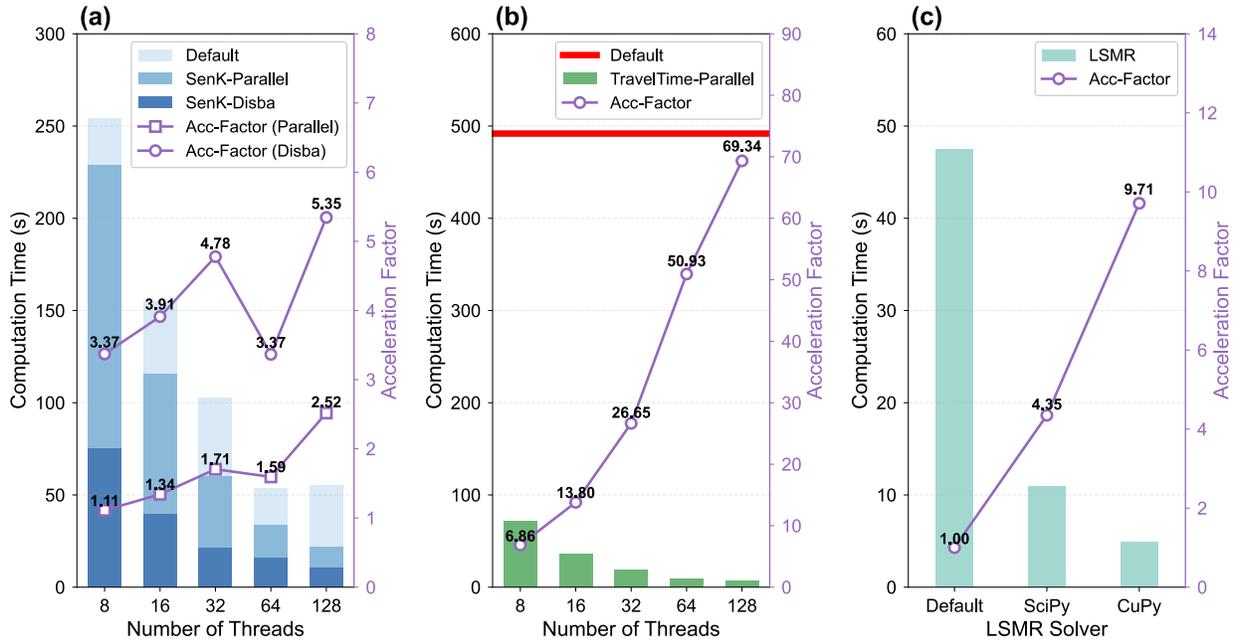

**Figure 5.** Computation time comparison of individual modules between the pDSurfTomo and DSurfTomo. "Default" denotes the native module of DSurfTomo. (a) Computation time and acceleration factor of the sensitivity kernel module under different thread counts. (b) Computation time and acceleration factor of the traveltime module under different thread counts. (c) Computational times and acceleration factors of the three least-squares solvers. Note that none of the evaluated solvers (native DSurfTomo-LSMR, SciPy-LSMR, and CuPy-LSMR) permit explicit user-level thread allocation; instead, thread concurrency is managed internally by their respective backend mathematical libraries.

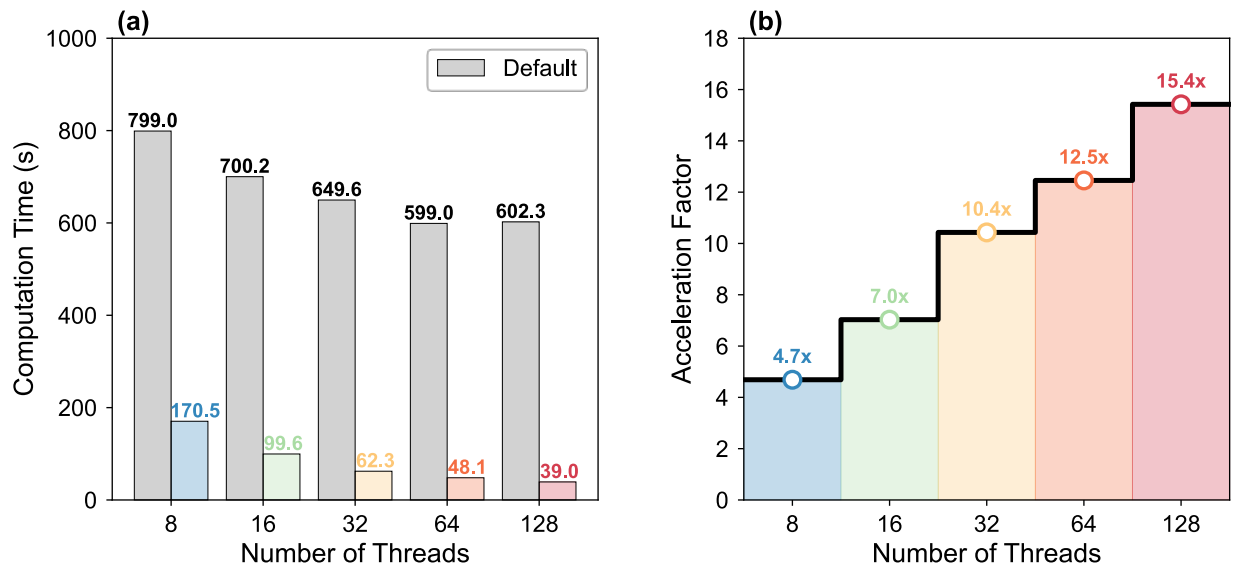

**Figure 6.** Comparison of the computation time per iteration between the pDSurfTomo and DSurfTomo. (a) Comparison times of both algorithms under different thread counts. Gray bars represent the Comparison time of DSurfTomo, while colored bars denote that of pDSurfTomo. (b) The corresponding acceleration factor achieved by pDSurfTomo under different thread counts.

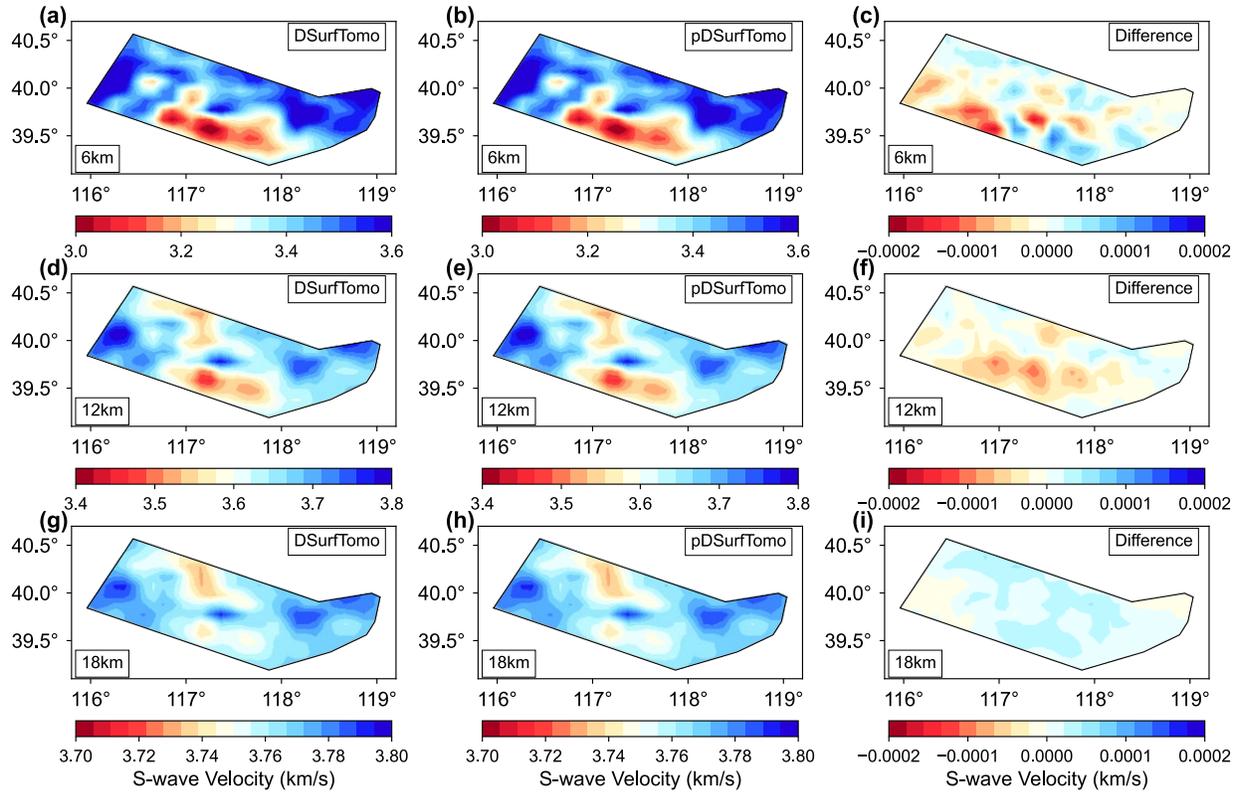

**Figure. 7.** Comparison of the inversion results by DSurfTomo and pDSurfTomo. (a)–(c) Horizontal slices of the inverted S-wave velocity model at the depth of 6 km by DSurfTomo, pDSurfTomo, and the difference between the two models, respectively. (d)–(f) are similar to (a)-(c) but for the depth of 12 km, and (g)–(i) are similar to (a)-(c) but for the depth of 18 km.

Supplement to

**pDSurfTomo: A High-Performance Parallel Computing Package for Direct Surface Wave Tomography**

Shaohang Zhu, Junlun Li*, Guoyi Chen, Hongjian Fang, Huajian Yao

**Description of the Supplemental Material**

Supplemental material for this paper includes figures illustrating four additional graphical user interfaces of pDSurfTomo.

**List of Supplemental Figure Captions**

**Figure S1.** Observation System Interface. This interface shows the spatial distribution of stations, surface wave ray path coverage, and inversion grid discretization, which allows users to intuitively verify the rationality of the observation system and adjust grid parameters interactively. Black triangles denote seismic stations, and gray dashed lines represent simplified surface wave ray paths.

**Figure S2.** MultiSlice Visualization Interface. This interface is designed for visualization and image export of the inversion results. Models from different iterations can be dynamically loaded and plotted with adjustable visualization parameters.

**Figure S3.** Orthogonal Slice Visualization. This interface shows orthogonal slices of the inverted velocity model at different iterations along the horizontal (Depth Slice), longitudinal (Lon Slice), and latitudinal (Lat Slice) planes.

**Figure S4.** SSH Interface. This interface is designed to establish a secure connection channel between the local client and the remote server. Once a connection is established, users can perform remote inversion tasks and visualize results solely through local graphical interaction

**Supplemental Figures**

**Figure S1**

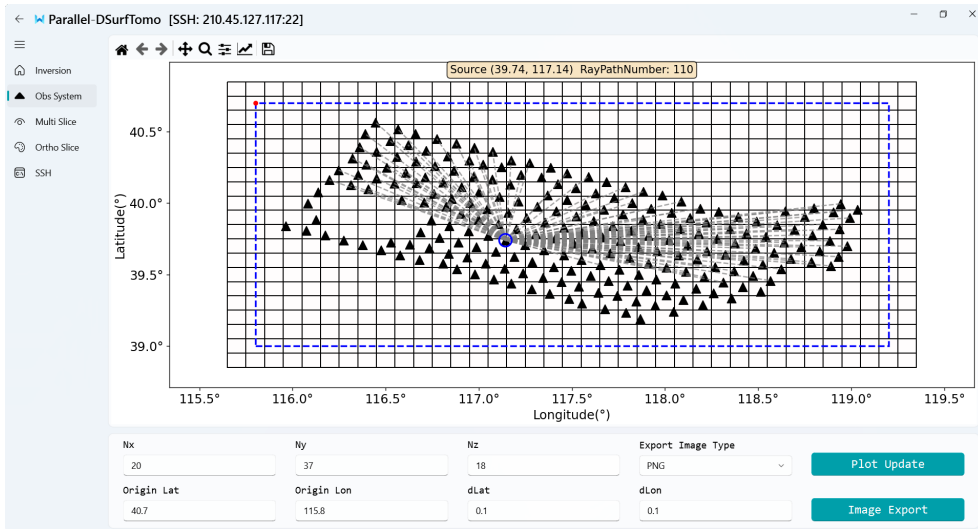

**Figure S2**

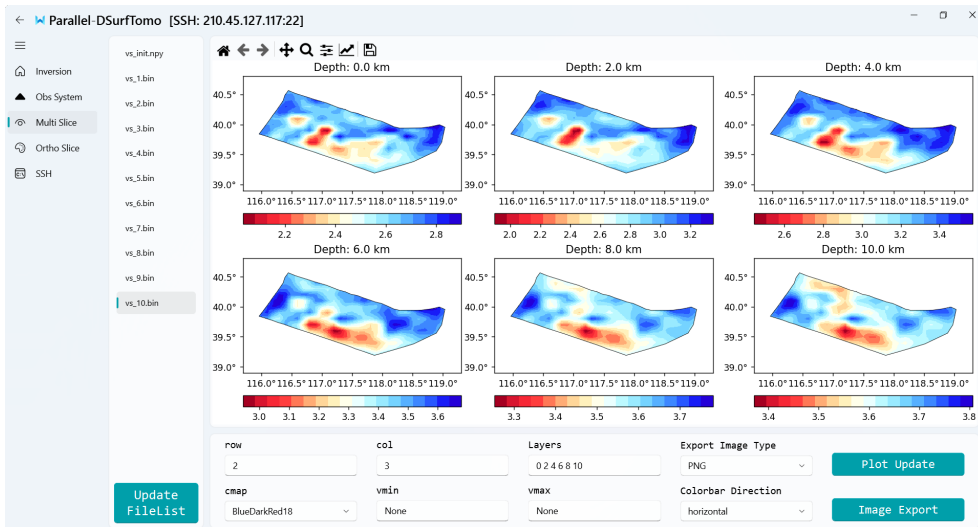

**Figure S3**

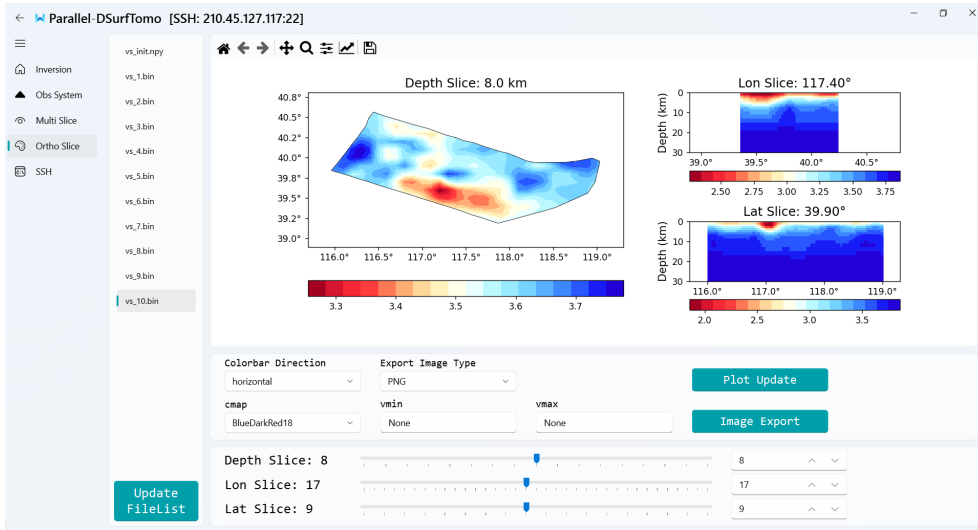

**Figure S4**

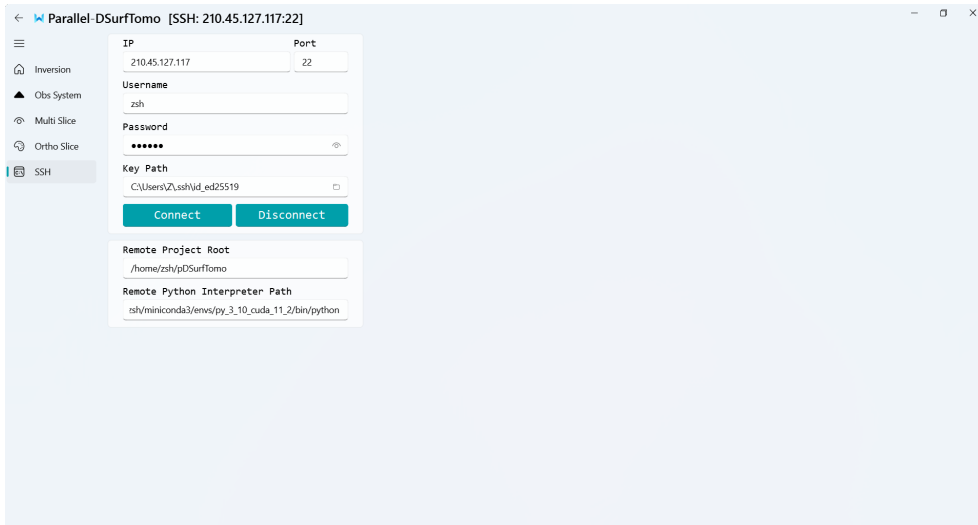